\documentstyle[epsfig]{aipproc}

\begin{document}
\title{Production of Beryllium and Boron by Spallation in Supernova Ejecta}
\author{Deepa  Majmudar$^1$, James H. Applegate$^2$}

\address{1:Dept. of Physics, 
Columbia University, 538 W. 120th Street, New York, NY 10027 \\
2:Dept. of Astronomy, 
Columbia University, 538 W. 120th Street, New York, NY 10027 \\
}

\maketitle
\vskip -3mm
\begin{abstract}

The abundances of beryllium and boron have been measured in halo stars of 
metallicities as low as [Fe/H] =-3.  The observations show that the ratios 
Be/Fe and B/Fe are independent of metallicity and approximately equal to
their solar values over the entire range of observed metallicity.  These 
observations are in contradiction with the predictions of simple models
of beryllium and boron production by spallation in the interstellar medium of 
a well mixed galaxy.   We propose that beryllium and boron are produced by
spallation in the ejecta of type II supernovae.  In our picture, 
protons and alpha particles are accelerated early in the supernova event
and irradiate the heavy elements in the ejecta long before the ejecta 
mixes with the interstellar medium.  We follow the propagation of the 
accelerated particles with a Monte-Carlo code and find that the energy
per spallation reaction is about 5 GeV for a variety of initial particle
spectra and ejecta compositions.  Reproducing the observed Be/Fe and B/Fe 
ratios requires roughly $3 \times 10^{47}$ ergs of accelerated protons 
and alphas.  This is much less than the $10^{51}$ ergs available in a 
supernova explosion.

\end{abstract}
\vskip -2mm

\section*{Introduction}

Spallation reactions involving protons or alpha particles colliding
with the nuclei of the abundant light elements carbon, nitrogen, and
oxygen have long been recognized as important or dominant contributors
to the production of the isotopes of lithium, beryllium, and boron (Reeves,
Fowler \& Hoyle 1970).   Evidence for this picture includes the fact that
the ratios of abundances of spallation products are equal to the ratios of
the production cross sections, the large overabundance of spallation 
products in cosmic rays, and the fact that the observed cosmic ray flux
irradiating a poulation I composition for the age of the galaxy produces
the observed spallation to CNO abundance ratio.  For a summary of these
arguments see the review by Reeves (1982).    

Spallation nucleosynthesis in the interstellar medium of a well mixed 
galaxy is very inefficient if the metallicity of the gas is low.  In the
simplest closed box model of galactic chemical evolution, the abundance of
spallation products is proportional to the square of the iron abundance at
low metallicity.   This prediction is contradicted by the observations of
beryllium and boron abundances in low metallicity stars 
(Duncan $et$ $al.$ 1992,
Gilmore $et$ $al.$ 1992, Boesgaard 1996) which show that the Be/Fe and B/Fe 
ratios are independent of metallicity and approximately equal to their solar
values for stars in the metallicity range -1$>$[Fe/H]$>$-3.

To account for these observations we propose that spallation
nucleosynthesis took place in the supernova event itself.  In our model, 
particles are accelerated and irradiate the CNO nuclei in the supernova
ejecta long before the ejecta mixes with the interstellar medium
of the early galaxy.   We follow the propagation of the accelerated particles
with a Monte-Carlo analysis and find that reproducing the observed 
spallation to iron ratio requires about $3 \times 10^{47}$ ergs to go
into accelerating particles.  We also find, not surprisingly, that we
produce the same isotopic ratios as are found in calculations which irradiate
a solar composition interstellar medium.

\section*{Model Description \& Results}


\begin{table}[b!]
\label{table1}
\caption{$E_{sp}$ (GeV) for various 
$E_{c}$ (MeV) and $n$,
calculated for a 25M$\odot$ supernova composition}
\begin{tabular}{lllll}
 $E_{c} (MeV)=$ &50 & 100 & 200 & 500  \\
\hline
n=5     &6.3    &4.3    &5.4    &9.6  \\
\hline
n=10    &4.3    &4.2    &4.9    &8.6   \\
\hline
\end{tabular}
\end{table}

A supernova releases a large
amount (10$^{51}$ergs) of energy as it explodes.
A fraction of this energy may go into accelerating 
particles to
very high energies.
The type II
supernova ejecta are rich in CNO nuclei which serve as
targets for spallation by energetic
protons and $\alpha$ particles that are
accelerated in the expanding ejecta.
 
We have modelled the spallation production of Li, Be and B in
supernova ejecta by writing a Monte-Carlo
simulation.
The program takes
as its input the spectrum of accelerated particles
(p, $\alpha$)
in supernova explosion and irradiates the ejecta
(CNO)
of specified composition and
density with these particles.
 
In the simulation, the accelerated particles
start out with initial energy as specified
by the spectrum
and have elastic, inelastic or spallation collisions
based on the probability of that collision according to
their relative cross sections and target abundance.
The energy dependent cross-sections
for these collisions are taken from
the compilations by Read \& Viola (1984) and Meyer (1971).
The main energy loss processes for these accelerated
particles are by spallation, ionization
and elastic/inelastic collisions.
Protons go through elastic and inelastic collisions with
other protons and He nuclei in the ejecta and spallation
reactions with CNO nuclei.
In the case of a proton-proton or a proton-$\alpha$
elastic collision,
the incident proton loses energy to the stationary
target and accelerates the target proton or $\alpha$
particle. This
creates a cascade of accelerated particles, each
of which in turn goes through a sequence of collisions.
The energy loss suffered by the incident proton in
a proton-proton elastic collision is
calculated kinematically using differential cross 
sections for elastic scattering (Meyer 1971).
The proton-proton inelastic collision, p + p $\rightarrow$
$\pi^{0}$ + p + p, produces $\pi^{0}$s which decay as
$\pi^{0}$ $\rightarrow$ $\gamma$ + $\gamma$, producing
gamma-ray flux. The remaining energy after $\pi^{0}$
production is shared between the two outgoing protons.
The proton-$\alpha$ inelastic collision produces secondary
particles (d,$^{3}$He) that are not relevent to our 
simulation
and the incident proton is assumed to lose all its energy. 
Spallation between protons and CNO produces one of the light
element isotopes 
according to their relative cross sections
(p + CNO $\rightarrow$
$^{6}$Li, $^{7}$Li, $^{9}$Be, $^{10}$B, $^{11}$B). 
Protons also  lose energy
by ionization in passing through the ejecta. These losses
begin to dominate at lower energies.
When the ionization energy loss is much greater ($\approx$
50 times)  
than the energy loss by other collisions
(spallation, elastic and inelastic), the proton is assumed
to lose all its energy by ionization and doesn't suffer
any more collisions.
Similar treatment is applied to accelerated $\alpha$
particles. In this case, the spallation reactions are
$\alpha$ + $\alpha$ $\rightarrow$ $^{6}$Li, $^{7}$Li and
$\alpha$ + CNO $\rightarrow$
$^{6}$Li, $^{7}$Li, $^{9}$Be, $^{10}$B, $^{11}$B. 


\begin{table}[b!]
\caption{Results  for $E_{c}$ = 100 MeV/nucleon, 
power law index $n$ = 5 calculated for various 
supernova masses}
\label{table2}
\begin{tabular}{lllll}
    & 15M$\odot$ &25M$\odot$ & 35M$\odot$  &Observed \\
\hline
B/Be      &16.4    &12.9    &12.5 &10-20$^{1}$ \\
\hline
$^{11}$B/$^{10}$B  &2.6    &2.5    &2.5  &4.0$^{2}$ \\
\hline
$^{7}$Li/$^{6}$Li   &1.5    &1.5    &1.6  &12.6$^{2}$ \\
\hline
$^{6}$Li/$^{9}$Be   &3.7    &3.0   &2.7  &3.7$^{2}$ \\
\hline
$N_{\pi0}$    & 2.0x10$^{47}$  & 6.5x10$^{46}$ & 4.4x10$^{46}$ & \\
\hline
$E_{sp}$ (GeV)  & 8.7  & 4.3 & 3.9 & \\
\hline
\end{tabular}
\tablenotetext{1. Duncan $et$ $al.$ (1992)  2. Cameron (1982)}
\end{table}

\begin{figure}[t!]
\centerline{\epsfig{file=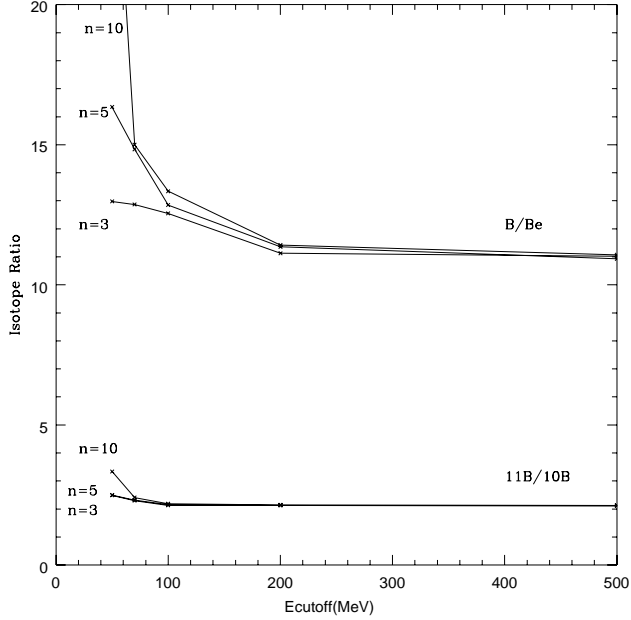,height=3.5in,width=3.5in}}
\vspace{10pt}
\caption{B/Be and $^{11}$B/$^{10}$B 
for various cutoff energies ($E_{c}$) and power law index $n$
using 25M$\odot$ supernova composition}
\label{fig:fps}
\end{figure}


\begin{table}[b!]
\caption{Number of $\pi^{0}$s produced
for various $E_{c}$ (MeV) and $n$,
calculated for 25M$\odot$ supernova}
\label{table3}
\begin{tabular}{lllll}
   & $E_{c}$=50 & 100 & 200 & 500  \\
\hline
n=5     &0    &6.5x10$^{46}$    &1.5x10$^{48}$    &2.8x10$^{49}$ \\
\hline
n=10    &0    &0                &3.4x10$^{46}$    &1.8x10$^{49}$   \\
\hline
\end{tabular}
\end{table}

The simulation is run for various accelerated
particle spectra and compositions of irradiated
material. The compositions of 15M$_{\odot}$, 25M$_{\odot}$ 
and
35M$_{\odot}$
supernova ejecta
(Weaver \& Woosley 1993)
are used as irradiated material distributed uniformly in
a sphere of radius of 10$^{15}$cm.
The source spectrum of accelerated particles 
is described
as   constant (flat spectrum) 
up to a certain cutoff energy $E_{c}$ and
at energies $E>E_{c}$, a power law decrease by 
index $n$.
We use various cutoff energies ($E_{c}$ = 50 MeV-500 MeV/nucleon)
and power law indices ($n$ = 2-10) and compare
the results.
 
We calculate the total number of light element
isotopes ($^{6}$Li, $^{7}$Li, $^{9}$Be, $^{10}$B, $^{11}$B) 
produced by spallation in the irradiated ejecta, the total 
number of elastic and inelastic collisions  and
the energy needed
per spallation $\left(E_{sp}\right)$. 

We find the
energy per spallation ($E_{sp}$) to be in the range of 1-10 GeV,
as shown in Table 1. The value of $E_{sp}$ is higher in the case of 
$E_{c}$ = 500 MeV/nucleon because the number of inelastic collisions
is dramatically increased as the accelerated particle spectrum 
extends well beyond the 280 MeV threshold for inelastic collisions. 

Using the  Solar abundances (Cameron 1982),
corrected for $^{7}$Li to account for its 
production in the big bang nucleosynthesis,
we get the total number of
spallations per $^{56}$Fe nucleus as 2.6 $\times$ 10$^{-5}.$
The total amount of $^{56}$Fe
ejected from a supernova is taken as 0.07M$_{\odot}$, 
as observed
in SN1987A (Erickson $et$ $al.$ 1988).
Therefore the total number of spallations per supernova
required for light element production is $3.9 \times 10^{49}$.
Using 5 GeV as the value of energy per spallation,
we get the total energy required as 
3 $\times$ 10$^{47}$ ergs.
A typical supernova releases about 
10$^{51}$ ergs, 
so
only a
small fraction of the total  energy 
needs to be directed towards spallation reactions. 

The variation of 
B/Be and $^{11}$B/$^{10}$B ratios with 
cutoff energy $E_{c}$
and power law index $n$ are shown in Fig. 1.
The isotopic ratios
$^{11}$B/$^{10}$B, $^{6}$Li/$^{9}$Be,
$^{7}$Li/$^{6}$Li
and
B/Be, the 
number of $\pi^{0}$s produced as a result of
p-p inelastic collision
and the energy required per spallation ($E_{sp}$), 
calculated for supernova masses 
of 15M$\odot$, 25M$\odot$ and  35M$\odot$
with accelerated particle spectrum as
$E_{c}$ = 100MeV/nucleon, index $n$ = 5 are shown 
in Table 2. The observed  isotopic ratios
are also shown for comparison. 
 
The B/Be ratio is recently observed in several Halo Dwarfs 
(Duncan $et$ $al.$ 1992,
Boesgaard 1996) and is found to be in the vicinity of 10, 
which is consistent
with our calculations. 
The Solar $^{11}$B/$^{10}$B ratio of 4 is higher than our
calculated value of 2.5. 
There may be other
sources of  $^{11}$B production, such as neutrino induced
nucleosynthesis in a type II supernova 
(Woosley $et$  $al.$ 1990), giving us the high $^{11}$B/$^{10}$B
ratio. 
 
The energy required per spallation decreases with increasing
supernova mass because the H/CNO ratio decreases with
increasing supernova mass and so there are more spallation
collisions and fewer elastic/inelastic collisions, thus
utilizing more of the energy for spallation. For the same
reason the number of $\pi^{0}$s produced decreases with
increasing supernova mass.

Table 3 shows
the number of $\pi^{0}$s produced for  various
cutoff energies $E_{c}$ and index $n$,
calculated for a 25M$\odot$ supernova composition.
The threshold for $\pi^{0}$ production is at 280 MeV,
so for certain accelerated particle spectra we get no pion
production. 
The $\pi^{0}$s subsequently decay producing two
$\gamma$ rays at energies centered around 70 MeV.

\end{document}